\documentstyle[epsfig,aps]{revtex}

\newcommand\Jou[4]{{#1} {\bf #2}, #3 (#4)}

\newcommand\NPA{{Nucl. Phys.} A}
\newcommand\NPB{{Nucl. Phys.} B}
\newcommand\PLB{{Phys. Lett.}  B}

\newcommand\PRL{Phys. Rev. Lett.}
\newcommand\PRC{{Phys. Rev.} C}
\newcommand\PRD{{Phys. Rev.} D}

\newcommand\ZPC{{Z. Phys.} C}

\newfam\BMath
\font\BMathL=cmmib10 
\font\BMathl=cmmib7
\font\BMathm=cmmib5
\textfont\BMath=\BMathL \scriptfont\BMath=\BMathl
\scriptscriptfont\BMath=\BMathm

\renewcommand\P{{\fam\BMath p}}

\newcommand\K{{\fam\BMath k}}

\renewcommand\a{\alpha}

\renewcommand\d{\delta}
\newcommand\D{\Delta}

\newcommand\g{\gamma}

\renewcommand\j{\Psi}

\newcommand\k{\kappa}

\renewcommand\L{\Lambda}
\newcommand\p{\pi}
\newcommand\q{\theta}
\newcommand\m{\mu}
\newcommand\n{\nu}

\newcommand\s{\sigma}
\renewcommand\t{\tau}
\newcommand\U{\Upsilon}

\newcommand\cf{{\cal F}}
\newcommand\ch{{\cal H}}
\newcommand\ci{{\cal I}}

\newcommand\co{{\cal O}}
\newcommand\cp{{\cal P}}

\newcommand\dd{\mbox{d}}

\renewcommand\exp{\mbox{\rm exp}}

\newcommand\ra{\rightarrow}

\newcommand\lra{\longrightarrow}

\newcommand\intpq{\frac{d^3 \P_q}{(2\pi)^3 2 p_q^0}}
\newcommand\intpqb{\frac{d^3 \P_{\bar q}}{(2\pi)^3 2 p_{\bar q}^0}}
\newcommand\intpqd{\frac{d^3 \P_{\bar q'}}{(2\pi)^3 2 p_{\bar q'}^0}}
\newcommand\intpg{\frac{d^3 \P_g}{(2\pi)^3 2 p_g^0}}
\newcommand\intkh{\frac{d^3 \K_{h}}{(2\pi)^3 2 k_{h}^0}}
\newcommand\intkho{\frac{d^3 \K_{h_1}}{(2\pi)^3 2 k_{h_1}^0}}
\newcommand\intkht{\frac{d^3 \K_{h_2}}{(2\pi)^3 2 k_{h_2}^0}}

\newcommand\ot{\otimes}

\newcommand\be{\begin{equation}}
\newcommand\ee{\end{equation}}
\newcommand\bea{\begin{eqnarray}}
\newcommand\eea{\end{eqnarray}}
\newcommand\ba{\begin{array}}
\newcommand\ea{\end{array}}
\newcommand\eref[1]{Eq.~(\ref{#1})}

\newcommand\fref[1]{Fig.~\ref{#1}}
\newcommand\tref[1]{Table~\ref{#1}}
\newcommand\bfi{\begin{figure}}
\newcommand\efi{\end{figure}}
\newcommand\bpi[1]{\begin{picture}#1}
\newcommand\epi{\end{picture}}

\newcommand{\ncom}{\newcommand}
\ncom{\lan}{\langle}
\ncom{\ran}{\rangle}
\ncom\fx{\!\!\!\!}

\ncom{\half}{\frac{1}{2}}
\ncom{\2}{\frac{1}{2}}
\ncom{\3}{\frac{1}{3}}
\ncom{\6}{\frac{1}{6}}
\ncom{\9}{\frac{1}{9}}
\ncom{\rtw}{\frac{1}{\sqrt{2}}}
\ncom{\rth}{\frac{1}{\sqrt{3}}}
\ncom{\rsi}{\frac{1}{\sqrt{6}}}
\ncom{\rtwth}{\sqrt{\frac{2}{3}}}

\begin{document}

%\begin{flushright}
%\footnotesize \sffamily NUC-MINN-99/11-T
%\end{flushright}

\draft
%\wideabs{

\title{Color Screening Effects on Hadronization 
in Relativistic Heavy Ion Collisions} 
%at the Relativistic Heavy Ion Collider}

\author{S.M.H. Wong}

\address{
School of Physics and Astronomy, University of Minnesota, Minneapolis, 
Minnesota 55455
}

\maketitle

\begin{abstract}

The effects of color screening on the hadronization of a 
parton plasma into a hadron gas are examined at the energies of the 
relativistic heavy ion collider. It is found to have the tendency 
to prevent hadronization and therefore delaying the conversion
of the partons into a hadron gas. Because of the continual expansion,
the resulting hadron gas number densities are lower when screening is
included. This should reduce the hadronic noise to genuine signals
of the quark-gluon plasma. In this sense, color screening is favorable
and should be included in numerical models. In any case, we advocate that 
numerical models should allow the confining forces and color screening to 
act on each other so as to undergo the phase transition in a natural way. 
Hadronization is also shown to seriously disrupt parton equilibration 
and is yet another reason why full parton chemical equilibration should 
not be expected. 

\end{abstract}

\pacs{PACS: 25.75.-q, 24.85.+p, 12.38.Mh, 12.38.Bx \hfill NUC-MINN-99/11-T}
%}

\section{Introduction}
\label{sec:intro}

The Relativistic Heavy Ion Collider (RHIC) at Brookhaven will be put into
operation in the imminent future. The primary aims of the 
experiments to be conducted at the new collider are to show that
deconfined matter can indeed exist under extreme conditions and also 
to uncover its properties. These are no easy tasks since the high
energy per nucleon combined with the multi-particle initial states
lead to the possibility of the many particles to undergo multiple
interactions. The resulting interactions are far more complicated
than what have been studied so far, not to mention the complex 
interactions that will take place in a region of sizable
spatial extent and temporal duration. These are novel aspects
for scattering experiments. This is especially true of the 
time evolution, which is usually not even under consideration
in $e^+ e^-$ annihilations or proton-antiproton collision
experiments due to the short duration of these processes. The final 
states or the end products are therefore the main concerns in these
cases. When facing many-body processes, numerical models are very 
essential. This is even more so in heavy ion collisions than in
$e^+ e^-$ annihilations or proton-antiproton collisions when 
the actual time evolution of the collisions becomes important. 
When modeling high energy nuclear interactions, it is necessary 
therefore to follow the partons through into the hadron phase
till the very end. The transition from partons into hadrons is
a considerable challenge. This is so because exactly
how quarks and gluons are bounded into hadrons is not well
understood. If one is only interested in the final states, one can
proceed as in \cite{fie&wolf,web1} for $e^+ e^-$ annihilation into
hadrons since the hadronization takes place after the primary
jet partons have a chance to evolve into parton showers 
through time-like parton branching. So this is to a certain extent
similar to the situation of that in a gas of partons. The major
difference is the latter exist in a much denser medium and there
is color screening whereas the former is essentially hadronization
in a vacuum. So one should keep well in mind of these differences
and include as much of these in any attempts at modeling
parton to hadron transitions in a QCD medium. As far as we can tell, 
there is no attempt in this direction in any realistic model.
In \cite{geig3}, attempt was made to incorporate hadronization in
the parton cascade model (PCM) \cite{geig1,geig2,geig&kap} 
but the hadronization process was essentially the same as that
used in $e^+ e^-$ in \cite{web1} and therefore medium effects
on hadron formation have been totally neglected. Note that one could
also study the transition to hadrons by using textbook type approaches
\cite{jbook} or by concentrating entirely on the macroscopic properties
such as bubble formation and growth \cite{ck1,ck2,ckkz}, but by doing so
most details that we are interested in will be hidden from view in
such coarse-grain approaches.

In our previous investigations in \cite{wong3}, we found that the average 
interaction strength in a parton plasma created in the wake of the initial 
hard collisions at RHIC or at the Large Hadron Collider at CERN tended to
increase monotonically in time or the interactions tended to become stronger 
and stronger with time. One of the consequences of this is that the screening 
mass, or more generally, the medium generated masses calculated at leading 
order also increases monotonically in time at least at RHIC and within the 
duration that we investigated, the screening length therefore decreases. 
In \fref{f:mass&len}, we plotted the time development of the square of 
the screening mass $m_D$ and medium generated quark mass $m_q$ and also the 
screening length $l_D$, all at leading order at RHIC which was calculated 
from the time evolution of the parton plasma in \cite{wong3}. They clearly 
show the above mentioned behavior\footnote{One might be alarmed that this 
mass did not decrease as the system cooled since it was well known that
the leading order term $m_D^{\text{LO}} \sim T^2$. However working only
at leading order in $\a_s$ meant that $m_D^{\text{LO}}$ also depended
on the factor $\ln^{-1} (T/{\L_{\overline{\text{MS}} }})$ (see \eref{eq:gsq}). 
The presence of the Landau pole is partly responsible for the behavior 
seen in \fref{f:mass&len}. The exact behavior depended on a number of 
factors such as the parton densities, the chemical composition at each 
moment, how fast the system is cooling etc.. At the energies of the 
Large Hadron Collider for example, the initial behavior is similar but 
the mass eventually went down with time because the contributing 
factors are different.}. The actual screening length at the end 
of the run when the temperature estimates fell to about 200 MeV 
was about 0.4 fm. This is comparable to the size of the most common 
hadrons. If the leading order results are anything to go by, this is not 
particularly good for hadronization if it really does occur at this time. 
To see what are the implications beyond leading order, we must turn to 
lattice calculations since perturbative calculation at the next-to-leading 
order required the introduction of the non-perturbative color magnetic mass 
as an infrared cutoff \cite{rebh}. In ref. \cite{kaj&etal1} the Debye 
screening mass $m_D$ was calculated for SU(N=3) on the lattice up to the term
at $\co (g^3)$ in the linear power dependence on $g$ but each term had a 
non-perturbative coefficient which may or may not have logarithmic, 
non-analytic $g$-dependence. It was found that in a large range of 
temperatures near and above the phase transition temperature $T_c$, it 
was several times larger than the leading order result $m_D^{\text{LO}}$. 
To be more precise, 
\be   m_D \; \simeq 3.3 \; m_D^{\text{LO}}   \; .
\ee
\bfi
\centerline{
\epsfig{figure=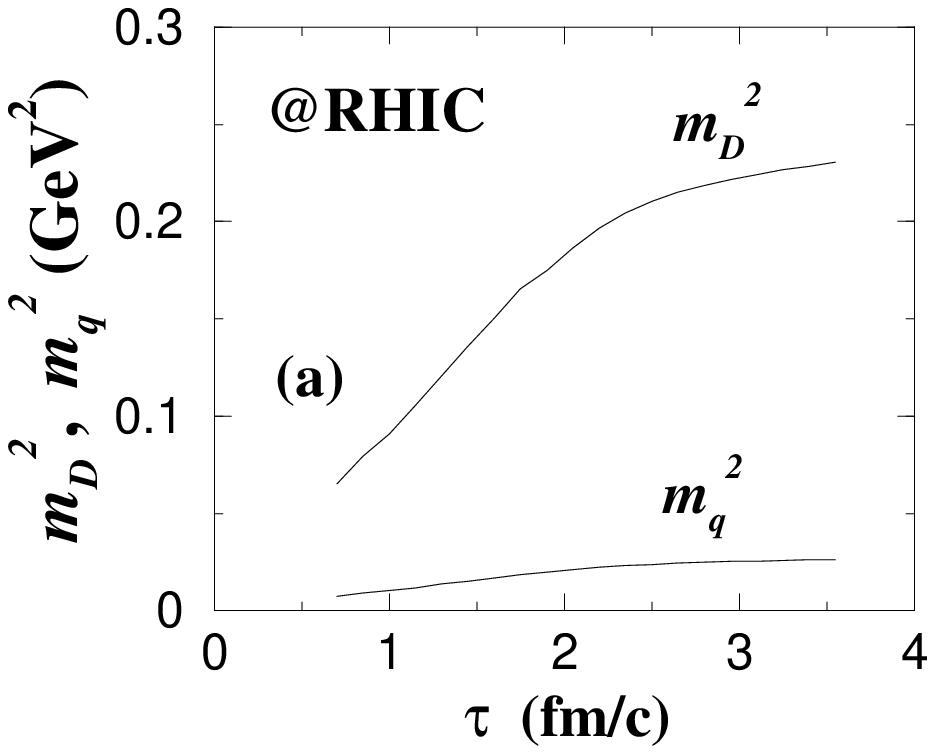,width=2.40in} \hspace{1.50cm}
\epsfig{figure=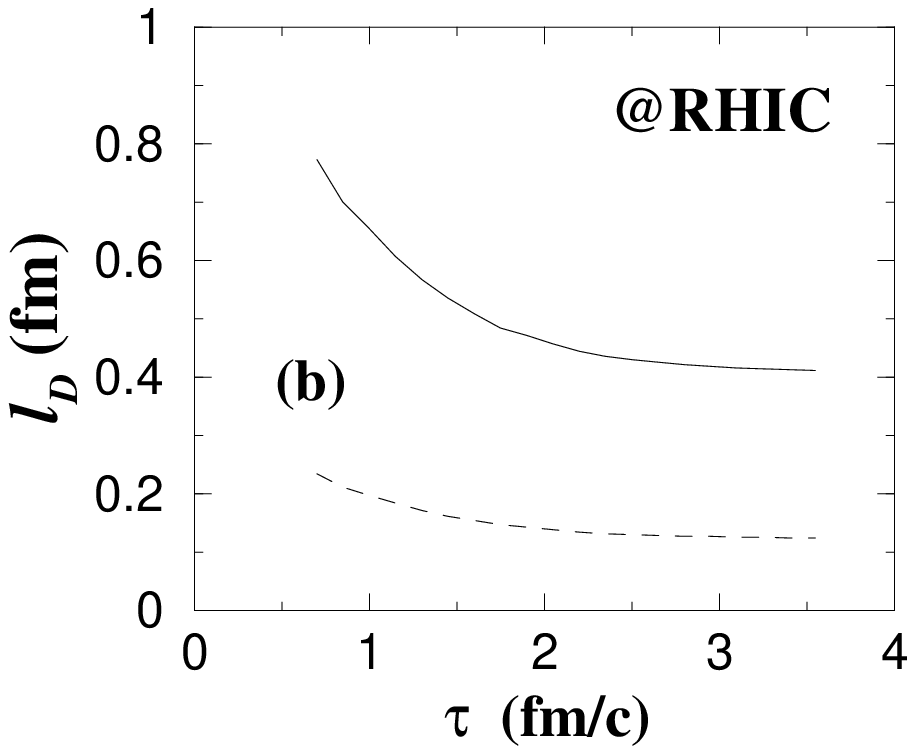,width=2.40in}
}
\caption{The time variation of (a) the medium masses at leading order and 
(b) the corresponding screening length from graph (a) at RHIC. The
dashed line in graph (b) is if we used a factor of 3.3 to get the $m_D$
result evaluated up to $\co(g^3)$ from lattice as discussed in the text.}
\label{f:mass&len}
\efi
From the equations 
\bea   m_D^{\text{LO}} & = & (N/3 +n_f/6)^{1/2} \, g\, T                       \\
       g^2 (\m)        & = & {{24 \p^2} \over {(11 N -2 n_f) 
                              \ln \m/{\L_{\overline{\text{MS}} }}} }
\label{eq:gsq}                                                                 \\ 
       \m \;\;         & = & 4 \p \; e^{-\g_E} \; T \;
           \exp \left ({{-3 c_m + 4 n_f \log 4} \over {22 N -4 n_f}} \right )  \\
       c_m \;          & = & {{10 N^2 + 2 n_f^2 +9 n_f/N} \over {6 N +3 n_f}}
\eea
given in \cite{kaj&etal1,kaj&etal2}, we can estimate the screening length 
at various temperatures. Using two flavors of quarks $n_f =2$ and 
$\L_{\overline{\text{MS}}} = 234$ MeV, $m_D^{\text{LO}} = 487.94 $ MeV at 
$T = 200$ MeV and $m_D^{\text{LO}} = 398.63 $ MeV at $T = 150$ MeV.
The screening length derived from the value of the non-perturbative $m_D$ 
up to $\co (g^3)$ would then be $l_D = 0.123$ fm at $T = 200$ MeV and 
$l_D = 0.150$ fm at $T = 150$ MeV. These are really small values when 
compared to even the smallest of hadrons. We must emphasize here once
again that although these results are up to $\co (g^3)$, they are not
perturbative because as we mentioned above already, the coefficient
at each power of $g$ is non-perturbative. In \fref{f:mass&len} (b), the 
leading order $m_D^{LO}$ in (a) was given a factor of 3.3 to arrive at 
the dashed curve to show the approximate small size of the would-be 
actual screening length if the screening mass was properly calculated 
non-perturbatively to beyond the leading order. Remember that the color 
screening property of a high energy and dense system of color charges is 
to reduce the color field exponentially within a distance of the screening 
length $l_D$. Thus color binding forces responsible for hadronizing, 
for example, the parton shower in $e^+ e^-$ annihilation would be weakened 
in a dense QCD parton medium. The conversion of partons into hadrons 
therefore cannot occur freely in heavy ion collisions. In fact, most 
hadrons cannot survive in an environment with the range of values of 
$l_D$ that were shown above. So if we impose by force the hadronization
time to be at any moment with $\t < 3.5$ fm without any regard to color
screening, it would not make any sense physically. The best way to
simulate hadronization, in our opinion, seems to be to let screening
and the confining forces act against each other and not to impose
a priori a specific temperature or density at which hadronization
must occur. The worse is of course to neglect color screening 
entirely. In this paper, we attempt to show the effects of color screened 
QCD medium has on hadronization and compare and contrast it with that 
without color screening or that occurring essentially in a vacuum. We 
will do this by letting the system decide by itself when to hadronize.

\section{Time evolution equations with hadronization}
\label{sec:t-evol}

\subsection{Basic Assumptions, Simplifications and Hadronization Mechanism}
\label{sec:ass_sim_mech}

Based on our time-evolution scheme developed in \cite{wong3,wong1,wong2}
for a parton plasma which we reported in \cite{wong4} and
applied to study several particle productions in \cite{wong5},
we now try to incorporate hadronization into the time evolution 
equations. In order to do this, we introduce the following simplifications.
\begin{itemize}

\item[i)]{The focus is in a region within a thin slice perpendicular
to the beam direction in the central region.}

\item[ii)]{This region is spatially homogeneous.}

\item[iii)]{There is only longitudinal Bjorken-like expansion within 
this region.}

\item[iv)]{There is boost invariance in the longitudinal or beam direction.}

\item[v)]{Surface effects will not be considered or the region of interest
will be the central core well away from the surface.}

\item[vi)]{Hadrons and resonances are formed by $q \,\bar q'$ only, 
where $q, q'= u, d$ or $s$.}

\item[vii)]{Resonances only undergo two-body decay.}

\item[viii)]{For hadrons, only $\p$'s and $K$'s are considered.}

\item[ix)]{Only orbital angular momentum $L \leq 1$ are allowed.}

\end{itemize}
The first four simplifications, i)--iv), are quite common and were used 
extensively in equilibration or hydrodynamics studies of the quark-gluon 
plasma and hydrodynamics studies of a hadron gas. Indeed, we relied heavily on 
them in our previous investigations of the time evolution of the parton plasma
\cite{wong3,wong1,wong2}. The v) is there because the surface regions are less 
interesting since color screening will be most important in the center where 
the parton density is highest. Near the surface hadrons will be
evaporating away from the system, whereas at the center hadronic bubbles will 
form instead. Although we do not explicitly use this type of macroscopic
language, our focus in this paper will be on the latter because color screening 
is less relevant near the surface. The vi) and the vii) were used in the studies 
of $e^+ e^-$ annihilations \cite{fie&wolf,web1} and we introduce viii) for 
reasons to be discussed below and also for the practical reason to reduce the 
time for computation. The last is purely there to avoid having to consider
too many possibilities.

Recalling that the main ingredients of our scheme 
consist of using a reduced form of the Boltzmann equation introduced 
by Baym \cite{baym} for the particle distribution $f_i$ and for the 
collision terms $C_i$, we used the relaxation
time approximation as well as constructing them explicitly from
parton interactions. The main time evolution equations for the 
partons thus appeared as 
\be {{\dd f_i^{\text{p}}} \over {\dd \t}} \, \Big |_{p_z \t =\text{constant}} 
    \; = - \; { {f_i^{\text{p}}-f^{\text{p}}_{i\; eq}}
            \over {\q_i^{\text{p}}} }
    \; =   \; C_i^{\text{p}} 
\label{eq:old_p_ev_eq}
\ee
where $i =g,q,\bar q$. The double construction of $C_i^{\text{p}}$ for
the parton collisions, hence the superscript p, is necessary to close the 
equations in order to be able to solve for the time-dependent parameters of 
the model. These will give us several time-dependent particle distributions 
$f^{\text{p}}_i$ which completely describe the time evolution of our partonic 
system. 

Here we very briefly recall the steps to arrive at the solution of 
\eref{eq:old_p_ev_eq}. This reduced Boltzmann equation has the advantage
that a partial solution in analytic form can be readily written down. 
That is 
\be f^{\text{p}}_i (p,\t) = f^{\text{p}}_{0\; i} (p,\t) \; \exp{(-x^{\text{p}})}
   + \int^{x^{\text{p}}}_0 d {x^{\text{p}}}' \; \exp{({x^{\text{p}}}'-x^{\text{p}})} \; 
     f^{\text{p}}_{\text{eq}\; i} (p,T^{\text{p}}_{\text{eq}\; i}(\t'),\t',\t)
\ee
where
\be x^{\text{p}}(\t) = \int^\t_{\t_0} d\t'/\q^{\text{p}}_i (\t')
\ee
and $f^{\text{p}}_{0\; i}$ is the initial distribution for parton 
species $i$ at $\t_0$. All there remains is to determine the parameters 
$T^{\text{p}}_{\text{eq}} (\t)$ and $\q^{\text{p}}(\t)$ for a complete 
solution. For this purpose, the perturbative construction of the 
collision terms was made. The two parameters were solved by converting 
any integral over $\t$ into a discrete sum as discussed in \cite{wong1}. 
For example, an integral function $G(\t)$ would be discretized as
\be G(\t)   = \int^\t_{\t_0} d\t' \tilde G(\t,\t') \;\; \Longrightarrow \;\;
    G(\t_n) = \sum^{n-1}_{i=0} \D \t \; \tilde G(\t_n,\t_i)  \; . 
\ee
Therefore provided that $\tilde G$ and any time-dependent parameters that it 
might have are known at all $\t < \t_n$, $G(\t_n)$ is determined completely.
When this method is applied to the distribution functions $f^{\text{p}}_i$,
they are determined at time $\t$ if the parameters $\q$ and $T_{\text{eq}}$
are known at all moments before $\t$. As a consequence, the medium masses,
energy densities, number densities, perturbatively constructed collision
terms etc. at any moment can be calculated by knowing the parameters of 
all earlier time-steps. The parameters of the next time-step are found by
equating the relaxation time approximation and the perturbatively
constructed collision terms and solving the resulting fourth degree
polynomials. These steps are for solving \eref{eq:old_p_ev_eq} for partons.
When hadrons are introduced in the later sections and given their own
transport equations, only the last step will be different. Instead of finding 
solutions to polynomials, we will be doing that to equations involving Bessel 
functions. 

To include hadronization in our model, one has to have a scheme. 
There are at least three different schemes that are being used in the studies 
of $e^+ e^-$ annihilations \cite{ell&stir&web}, namely string fragmentation,
independent parton fragmentation and parton clustering. Since we
prefer the parton language, we will not choose the first scheme.
Of the remaining two, independent parton fragmentation \cite{fie&fey1,fie&fey2}
tends to always end with a parton so this scheme by itself cannot convert a parton 
gas completely into hadrons. The more suitable scheme therefore is the third
scheme of parton clustering into hadrons \cite{fie&wolf,web1}. It may be that 
the combination of the second and the third is the more correct parton
based picture but we will use only one scheme for simplicity. 

To form hadrons via clustering, any number of partons can in principle 
participate in any one single cluster provided that they form a color singlet. 
It has to be said that the binding of a number of partons into a color 
singlet cluster does not directly result in the observed hadrons, rather 
they will appear as the decay products of the first formed cluster. 
For hadrons at least and it is reasonable to assume the same to be
true for the first formed resonance cluster, the probability of a hadron in a 
state with many constituents is much less than if it is in the valence 
state. So instead of allowing any number of partons to form any one
cluster, we will only allow two, that is a quark and an antiquark to
keep the scheme simple. This is the reasoning behind simplification vi) above.
There is, however, an obvious obstacle to this simple scheme. 
The parton plasma is well known to be gluon dominated
because of the small $x$ growth of the gluon distribution in the
nucleon. Barring the possibility of forming any glueball mesons,
the gluons must be converted into fermions somehow to fit in with
our simple hadronization scheme. Of course there is always the
perturbative conversion of gluons into quark-antiquark pairs but
it is a too inefficient process. Quite similarly in the studies
or modeling of $e^+ e^-$ annihilations, it was also found that
the time-like branching of gluon into quark-antiquark was much
less probable than that of branching into gluon, so it was 
necessary to resort to some non-perturbative mechanism to
ensure that the parton branching always ends in a quark-antiquark pair. 
A non-perturbative gluon splitting mechanism into quark-antiquark
was introduced in \cite{fie&wolf} for this purpose. Here we will
therefore also introduce such a process. However, the conversion 
mechanism that we will use below in the next section can only
be similar in essence, but not in detail to that used in $e^+ e^-$ 
annihilation. This is so because the parton shower created from
the parton branching consists entirely of off-shell time-like partons
whereas the partons in the parton plasma are essentially massless
and on-shell. We will discuss this further below.

\subsection{Parton-Hadron Conversion}
\label{sec:part-had}

By bringing in all the afore-mentioned mechanisms to complete our
simple hadronization scheme, our time evolution equations for
partons given in \eref{eq:old_p_ev_eq} become
\be {{\dd f_i^{\text{p}}} \over {\dd \t}} \, \Big |_{p_z \t =\text{constant}} 
    \; = - \; { {f_i^{\text{p}}-f^{\text{p}}_{i\; eq}}
            \over {\q_i^{\text{p}}} }
    \; =   \; C_i^{\text{p}} + C^{\text{p}}_{i\; g\ra q} 
            + C^{\text{p}}_{i\; p\ra h} \; .
\label{eq:new_p_ev_eq}
\ee
We have not included the possibility of parton-hadron interactions
and will not do so in this paper.
By introducing hadrons into the system, one time evolution equation
must be included for each hadron type. Using the same relaxation
time approximation for the partons as for hadrons, we have the
following time evolution equations for the hadrons
\be {{\dd f_i^{\text{h}}} \over {\dd \t}} \, \Big |_{p_z \t =\text{constant}} 
    \; = - \; { {f_i^{\text{h}}-f^{\text{h}}_{i\; eq}}
            \over {\q_i^{\text{h}}} }
    \; =   \; C^{\text{h}}_{i\;p \ra h} \; ,
\label{eq:new_h_ev_eq}
\ee
where $i =\p, K$ since we only include pions and kaons, and the 
interactions amongst hadrons are also not considered because they are not
our main focus here. The form of the parton collision terms 
$C_i^{\text{p}}$ can be found in \cite{wong1,wong2}. The new terms 
$C^{\text{p}}_{i\; g\ra q}$, $C^{\text{p}}_{i\; p\ra h}$ and 
$C^{\text{h}}_{i\;p \ra h}$ introduced here will be given and explained
below. 

Considering only three flavors and allowing only $q \bar q'$ to
form color singlet resonance clusters, the resulting clusters
can have their constituents in different spin and angular momentum
combinations. This together with the flavor of the fermion constituents
put them into definite parity, C-parity, angular momentum and
isospin state. These must be conserved in the subsequent strong decay
into observable hadrons. Because of simplification vii) and viii) and
the conservation laws, with only two pseudo-scalars in the decay
products means that not all possibilities of parity $\cp$, C-parity, spin $S$
and orbital $L$ or total $J$ angular momentum of the resonances are possible. 
Using the $J^{\cp (C)}$ notation, where the $C$ only applies when
we have flavor neutrality or equivalently when the resonance is an eigenstate 
of charge conjugation operator $C$, resonance clusters can only 
be found in any of the $1^{-(-)}$, $0^{+(+)}$ or $2^{+(+)}$ under the 
restrictions that we imposed.

For any given $q\bar q'$, if the kinematics allows it, we will assume
that there are no preferences in the binding of the pair into each of the
three possible cluster states given above. Thus for a number of $q\bar q'$
at any moment, an equal portion of them at that moment will choose to form 
each of the possible resonances. Likewise, we use the same assumption for the
subsequent decays of these resonances into each possible channel. 
For all kinematically allowed channels, equal number of any given kind of
resonances at any moment will choose to go into each of the possible 
hadron pairs, although the conversion rate for different channels may
not be the same due to other factors such as mass differences and
the available phase space.   

In the decay process, because we have assumed that both the resonance
clusters and the mesons are formed of $q\bar q'$, a pair of quark-antiquark
has to be created. We allow $u, d$ or $s$ flavor to have equal chance
to be created. One may object to this equal probability even for
the $s$ quark because it is so much heavier than the light quarks.
The reason that the same probability is assigned even to the strange
flavor is because we prefer to let phase space takes over so that
the smaller number of kaons in the final hadron abundance would
be due entirely to the available phase space. The latter has a major
role to play in the final particle abundances as was discovered
by Hagedorn in the studies of large angle p-p collision in the 50's.
This has, following various events, led to the construction of the 
statistical bootstrap model \cite{hag1,hag2}. 
This is the basic reason why we choose to let phase space decide the 
hadron yields rather than introducing some ad hoc, uneven probability 
for any flavor of $q\bar q$ pair creation in the decay process. 
As to the rate of conversion of the partons into hadrons, it is
controlled to a definite extent by isospin constraint. 
In \tref{t:chance1}, \tref{t:chance2} and \tref{t:chance3}, we give
the relative probabilities $\cf_{q\bar q'}^{hh'}$ of any given $q\bar q'$ with 
invariant mass $M$ to go through the three possible resonance cluster states 
and to end up as any of the possible final hadron pairs $hh'$. Not all 
possibilities of fermion pairs are given, those not given have the same 
relative probabilities as their charge conjugated pairs. 

\noindent\begin{minipage}[h]{\columnwidth}
\begin{table}
\caption{Relative probabilities $\cf_{q\bar q'}^{hh'}$ of any $q \bar q'$ 
to go through different clustering and into the decay channels indicated in 
parenthesis in the cluster invariant mass range $2\; m_\p \leq M < m_\p + m_K$.}
\label{t:chance1}
\begin{tabular}{@{\hspace{.2in}}llll@{\hspace{.2in}}}
$q\bar q'\;\big\backslash\;\cf_{q\bar q'}^{hh'} \big\backslash$ 
{\scriptsize cluster\ $J^{\cp (C)}$} 
             & $1^{-(-)}$         & $0^{+(+)}$         & $2^{+(+)}$           \\ \hline
             &                    &                    &                      \\
$u\;\bar d$  & $1\; (\p^0\;\p^+)$ & $\emptyset$        & $\emptyset$          \\
$u\;\bar u$  & $\3\;(\p^+\;\p^-)$ & $\6\;(\p^+\;\p^-)$ , $\6\;(\p^0\;\p^0)$
             & $\6\;(\p^+\;\p^-)$ , $\6\;(\p^0\;\p^0)$                        \\   
$d\;\bar d$  & $\3\;(\p^+\;\p^-)$ & $\6\;(\p^+\;\p^-)$ , $\6\;(\p^0\;\p^0)$
             & $\6\;(\p^+\;\p^-)$ , $\6\;(\p^0\;\p^0)$                        \\   
$u\;\bar s$  & $\emptyset$        & $\emptyset$        & $\emptyset$          \\
$d\;\bar s$  & $\emptyset$        & $\emptyset$        & $\emptyset$          \\
$s\;\bar s$  & $\emptyset$        & $\emptyset$        & $\emptyset$          \\
\end{tabular}
\end{table}
\end{minipage}
%\vspace{0.012cm}
%\newpage

If the QCD Lagrangian for three flavors could be rewritten into an
effective one which contains terms that describe binding of,
for example, $q\bar q'$, $q\bar q' g$, $q\bar q' g g$, $\dots$ into
hadrons and this Lagrangian is flavor blind, then the rate of our 
simplified parton-hadron conversion could be written with only three unknown 
transition matrix functions, which in general would have some mass dimensions. 
These functions would describe the isospin zero-to-zero, one-to-one and 
half-to-half confinement transitions. In \tref{t:trans_matr}, we give
the isospin coefficients $\U_{q\bar q'}^{hh'}$ and the dependence of 
each transition on the three functions which are written as 
$\lan I=0|\ch|I=0 \ran$, $\lan I=1|\ch|I=1 \ran$ and 
$\lan I=1/2|\ch|I=1/2 \ran$. 

\noindent\begin{minipage}[h]{\columnwidth}
\begin{table}
\caption{Relative probabilities $\cf_{q\bar q'}^{hh'}$ of any $q \bar q'$ 
to go through different clustering and into the decay channels indicated in 
parenthesis in the cluster invariant mass range $m_\p + m_K \leq M < 2\; m_K $.}
\label{t:chance2}
\begin{tabular}{@{\hspace{.2in}}llll@{\hspace{.2in}}}
$q\bar q'\;\big\backslash \;\cf_{q\bar q'}^{hh'} \big\backslash$ 
{\scriptsize cluster\ $J^{\cp (C)}$} 
             & $1^{-(-)}$         & $0^{+(+)}$         & $2^{+(+)}$           \\ \hline
             &                    &                    &                      \\
$u\;\bar d$  & $1\; (\p^0\;\p^+)$ & $\emptyset$        & $\emptyset$          \\
$u\;\bar u$  & $\3\;(\p^+\;\p^-)$ & $\6\;(\p^+\;\p^-)$ , $\6\;(\p^0\;\p^0)$
             & $\6\;(\p^+\;\p^-)$ , $\6\;(\p^0\;\p^0)$                        \\   
$d\;\bar d$  & $\3\;(\p^+\;\p^-)$ & $\6\;(\p^+\;\p^-)$ , $\6\;(\p^0\;\p^0)$
             & $\6\;(\p^+\;\p^-)$ , $\6\;(\p^0\;\p^0)$                        \\   
$u\;\bar s$  & $\6\;(\p^0\;K^+) $ , $\6\;(\p^+\;K^0) $ 
             & $\6\;(\p^0\;K^+) $ , $\6\;(\p^+\;K^0) $ 
             & $\6\;(\p^0\;K^+) $ , $\6\;(\p^+\;K^0) $                        \\   
$d\;\bar s$  & $\6\;(\p^0\;K^0) $ , $\6\;(\p^-\;K^+) $ 
             & $\6\;(\p^0\;K^0) $ , $\6\;(\p^-\;K^+) $ 
             & $\6\;(\p^0\;K^0) $ , $\6\;(\p^-\;K^+) $                        \\   
$s\;\bar s$  & $\emptyset$        & $\emptyset$        & $\emptyset$          \\
\end{tabular}
%\end{table}
\vspace{0.20cm}
%\begin{table}
\caption{Relative probabilities $\cf_{q\bar q'}^{hh'}$ of any $q \bar q'$ 
to go through different clustering and into the decay channels indicated in 
parenthesis in the cluster invariant mass range $2\; m_K \leq M $.}
\label{t:chance3}
\begin{tabular}{@{\hspace{.2in}}llll@{\hspace{.2in}}}
$q\bar q'\;\big\backslash \;\cf_{q\bar q'}^{hh'} \big\backslash$ 
{\scriptsize cluster\ $J^{\cp (C)}$} 
             & $1^{-(-)}$         & $0^{+(+)}$         & $2^{+(+)}$           \\ \hline
             &                    &                    &                      \\
$u\;\bar d$  & $\6\;(\p^0\;\p^+)$ , $\6\;(K^+\;\bar {K^0}) $ 
             & $\3\;(K^+\;\bar K^0)$  & $\3\;(K^+\;\bar K^0)$                 \\ 
$u\;\bar u$  & $\6\;(\p^+\;\p^-)$ , $\6\;(K^+\;K^-)$ 
             & $\9\;(\p^+\;\p^-)$ , $\9\;(\p^0\;\p^0)$ ,
             & $\9\;(\p^+\;\p^-)$ , $\9\;(\p^0\;\p^0)$ ,                      \\
             &                    & $\9\;(K^+\;K^-)$   & $\9\;(K^+\;K^-)$     \\
$d\;\bar d$  & $\6\;(\p^+\;\p^-)$ , $\6\;(K^0\;\bar K^0)$ 
             & $\9\;(\p^+\;\p^-)$ , $\9\;(\p^0\;\p^0)$ ,
             & $\9\;(\p^+\;\p^-)$ , $\9\;(\p^0\;\p^0)$ ,                      \\
             &                    & $\9\;(K^0\;\bar K^0)$   
                                  & $\9\;(K^0\;\bar K^0)$                     \\
$u\;\bar s$  & $\6\;(\p^0\;K^+) $ , $\6\;(\p^+\;K^0) $ 
             & $\6\;(\p^0\;K^+) $ , $\6\;(\p^+\;K^0) $ 
             & $\6\;(\p^0\;K^+) $ , $\6\;(\p^+\;K^0) $                        \\   
$d\;\bar s$  & $\6\;(\p^0\;K^0) $ , $\6\;(\p^-\;K^+) $ 
             & $\6\;(\p^0\;K^0) $ , $\6\;(\p^-\;K^+) $ 
             & $\6\;(\p^0\;K^0) $ , $\6\;(\p^-\;K^+) $                        \\   
$s\;\bar s$  & $\6\;(K^+\;K^-)$   , $\6\;(K^0\;\bar K^0)$   
             & $\6\;(K^+\;K^-)$   , $\6\;(K^0\;\bar K^0)$   
             & $\6\;(K^+\;K^-)$   , $\6\;(K^0\;\bar K^0)$                     \\
\end{tabular}
%\end{table}
%\end{minipage}
\vspace{0.20cm}
%\noindent\begin{minipage}[h]{\columnwidth}
%\begin{table}
\caption{The dependence of each allowed parton-hadron transition on
the three transition matrix functions and the associated isospin factor
$\U_{q\bar q'}^{hh'}$ for the decay channels shown in parenthesis.}
\label{t:trans_matr}
\begin{tabular}{@{\hspace{.2in}}llll@{\hspace{.2in}}} 
$q\bar q'\;\big\backslash\; \U_{q\bar q'}^{hh'}\; \big\backslash$ 
${\text{isospin transition} \atop \text{dependence}}$
             & $\lan I=0|\ch|I=0 \ran$              & $\lan I=1|\ch|I=1 \ran$    
             & $\lan I=1/2|\ch|I=1/2 \ran$                               \\ \hline
             &                 &                    &                    \\
$u\;\bar d$  & $\emptyset$     & $1(\p^0\;\p^+)$ , $1(K^+\;\bar K^0)$ 
             & $\emptyset$                                               \\            
$u\;\bar u$  & $\rth(\p^+\;\p^-)$ , $\rsi(\p^0\;\p^0)$ ,
             & $\rtw(\p^+\;\p^-)$ , $\2(K^+\;K^-)$  & $\emptyset$        \\
             & $\2(K^+\;K^-)$  &                    &                    \\ 
$d\;\bar d$  & $\rth(\p^+\;\p^-)$ , $\rsi(\p^0\;\p^0)$ ,
             & $\rtw(\p^+\;\p^-)$ , $\2(K^0\;\bar K^0)$  & $\emptyset$   \\
             & $\2(K^0\;\bar K^0)$  &                    &               \\ 
$u\;\bar s$  & $\emptyset$     & $\emptyset$        
             & $\rth(\p^0\;K^+)$ , $\rtwth(\p^+\;K^0)$                   \\
$d\;\bar s$  & $\emptyset$     & $\emptyset$        
             & $\rth(\p^0\;K^0)$ , $\rtwth(\p^-\;K^+)$                   \\
$s\;\bar s$  & $\rtw(K^+\;K^-)$  , $\rtw(K^0\;\bar K^0)$
             & $\emptyset$     & $\emptyset$                             \\ 
\end{tabular}
\end{table}
\end{minipage}

Then the collision terms for a quark of flavor $q$ to undergo a transition can 
be written as 
\bea  C^{\text{p}}_{q\; p\ra h} 
      &=&-\frac{1}{2 p_q^0} \; S(\a_s) \; \sum_{\bar q', h_1, h_2} 
          \int \intpqd \intkho \intkht
          (2\p)^4 \d^{(4)} (p_q +p_{\bar q'} -k_{h_1} -k_{h_2})   \nonumber \\
      & & \mbox{\hspace{3.0cm}} \times
          f_q f_{\bar q'} (1+f_{h_1}) (1+f_{h_2}) \; \cf_{q\bar q'}^{h_1h_2} \;
          \big |\sum_{I \in \ci} \U_{q\bar q'}^{h_1h_2} \lan I |\ch| I \ran \big |^2
          \; ,                                                              \\
      C^{\text{p}}_{g\; p\ra h} &=& 0   \; ,                                        
\eea
where $\ci = (I_q\ot I_{\bar q'}) \cap (I_{h_1}\ot I_{h_2}) $.
The hadron collision terms for the same conversion would be
\bea  C^{\text{h}}_{h_1\; p\ra h} 
      &=& \frac{2 N}{2 k_{h_1}^0} \; S(\a_s) \; \sum_{q, \bar q', h_2} 
          \int \intpq \intpqd \intkht
          (2\p)^4 \d^{(4)} (p_q +p_{\bar q'} -k_{h_1} -k_{h_2})   \nonumber \\
      & & \mbox{\hspace{3.0cm}} \times
          f_q f_{\bar q'} (1+f_{h_1}) (1+f_{h_2}) \; \cf_{q\bar q'}^{h_1h_2} \;
          \big |\sum_{I \in \ci} \U_{q\bar q'}^{h_1h_2} \lan I |\ch| I \ran \big |^2
          \; .
\eea
A factor $S(\a_s)$ is inserted to progressively switch on these confining
terms depending on the size of $\a_s$. We have chosen the form 
$S(\a_s) =\exp\{-(1-\a_s)/\s\}$ with $\s = 0.05$ to give a fairly sharp rise 
when $\a_s$ became large. Note that we have combined the two-step process 
that were the color singlet cluster formation and the subsequent decay into 
a one-step process so that a separate time evolution equation for the resonance 
clusters would not have to be introduced. The transition matrix elements now 
describe the two-in-one combined transition. These resulting terms would 
convert quarks and antiquarks into hadrons, but they by themselves would not be 
sufficient for hadronization because there are far more gluons than fermions 
in the parton plasma. To ensure the complete conversion into hadrons, another 
new term has to be introduced into the time evolution equations. 

As we mentioned above, by using a simple hadronization scheme,
the more numerous and dominant gluons of the parton plasma have
to be converted into quark-antiquark pairs and the conversion
process will have to be much more efficient than the leading
order perturbative process. The most straight forward way around
this is to use a non-perturbative gluon splitting mechanism
such as the one used in \cite{web1}, which always ensures that the
chains of coherent angular ordered parton branching end up as
quark-antiquark pairs in the modeling of $e^+ e^-$ annihilations.
Since our parton plasma has time to come on-shell and to interact,
any coherence would have long been destroyed in the thermalization,
as such we cannot use the detailed techniques of the gluon splitting
mechanism of \cite{web1}. Nevertheless it is reasonable
to preserve, if not in details, the idea of this mechanism during
hadronization. Since confinement is a localized microscopic process 
\cite{ama&ven,bas&cia&marc,mar&tre&ven}
which should not be affected by the global state of the parton system
whether it is in thermal equilibrium or not. The latter will, however,
affect to a certain extent the momentum distributions of the final 
hadrons. We will therefore have to introduce a new term 
$C^{\text{p}}_{i\; g\ra q}$ into the equations that fulfills the
purpose of a more efficient gluon-to-fermion pair conversion mechanism. 
We write
\bea C^{\text{p}}_{g\; g\ra q} &=& -\frac{1}{2 p^0_g}    \;S(\a_s) \;\sum_{q=u,d,s} 
         \int \intpq \intpqb (2\p)^4 \d^{(4)} (p_g -p_q -p_{\bar q}) \nonumber \\
     & & \mbox{\hspace{3.0cm}} \times
         f_g (1-f_q) (1-f_{\bar q}) \;  |\lan q\bar q|\ch| g\ran |^2        \\
     C^{\text{p}}_{q\; g\ra q} &=&  \frac{\n_g}{2 p^0_q} \;\frac{S(\a_s)}{2N} \;
         \int \intpg \intpqb (2\p)^4 \d^{(4)} (p_g -p_q -p_{\bar q}) \nonumber \\
     & & \mbox{\hspace{3.0cm}} \times
         f_g (1-f_q) (1-f_{\bar q}) \;  |\lan q\bar q|\ch| g\ran |^2 \; .
\eea
$\n_g$ here is the gluon multiplicity.
Strictly speaking massless gluons cannot split into massless quark-antiquark 
pairs, but considering that gluons in a QCD medium would in any case 
acquired a finite medium mass or that confinement would force the gluons
to go off-shell, such a minor restriction will be relaxed here.
Actually from the physical point of view it is not minor, but the main point is to 
ensure the conversion of $g \lra q\bar q$ in terms of the transfer of their 
numbers and energy-momentum into those of the fermion pairs and viewing from 
this angle this restriction is less important.

\subsection{Hadronization in a Color Screened Medium}
\label{sec:cs_had}

In the previous subsection, a simple scheme to achieve parton-hadron 
conversion was given. Although the basic conversion mechanism is there
and the state the partons are in when the conversion starts to take
place is different from other more common situations in which hadronization
schemes are applied, it is actually not much different from that occurring in a
vacuum. As mentioned in the introduction, medium effects are expected. 
The most important one in connection with hadronization should 
be the expected effects from color screening. It is one of the most important
because it is the core mechanism responsible for the much discussed 
signature proposed for the quark-gluon plasma --- $J/\psi$ suppression first 
proposed in \cite{ms} --- that should signal the presence of a deconfined 
medium. For this reason, any parton based model that does not incorporate 
medium effects cannot estimate dynamically the magnitude of charmonium
suppression due to color screening. We will include it in a way 
that captures the essential physics of color screening. The basic
reasoning of our method is as follows. Any hadrons must have a certain
physical size, which can be thought of as the internal separation
of the quark-antiquark pair in the case of a meson because of the 
intrinsic internal motion of these constituents. Since the pair is under
normal circumstances kept together by the confining color force, in the
presence of a color screened medium, the confining force should no longer
be able to keep the pair together if their mutual separation becomes equal or
larger than the screening distance over which the color field strength
dropped exponentially. Based on this simple fact, we take the spatial
separation of the pair to have a Gaussian like distribution
\be     F_h(b) = \frac{2}{a_h \p} \; \exp \; 
               {\left (-\frac{b^2}{a^2_h \p} \right)}  \; ,
\ee 
which is normalized according to 
\be     \int^\infty_0 \dd b \; F_h(b) = 1      \; .
\ee
The average size of the hadron $h$ would then be 
\be     a_h  = \lan \;b\; \ran = \int^\infty_0 \dd b \; b \; F_h(b)   \; .
\ee
In fact, the form of the distribution would likely be different for different 
orbital angular momentum states but such fine details are not warranted here.
In any case, simplification viii) permits only $\p$'s and $K$'s which are
S-wave hadrons\footnote{For hadrons with zero orbital angular momentum, 
another reasonable, possible form for the distribution would be 
$F_h(b)= a_h^{-1}\;\exp (-b/a_h)$. The resulting probability is however
quite similar to \eref{eq:problt} and hence will not affect the qualitative
picture which we will try to extract below.}.
The probability of finding a hadron with spatial separation less than 
the screening length $l_D(\t)$ of the medium at time $\t$ would then be
\be     P (b < l_D(\t)) = \int^{l_D(\t)}_0 \dd b \; F_h(b) 
                        = \text{Erf} \; (l_D(\t)/a_h \sqrt{\p})       \; .
\label{eq:problt}
\ee  
Erf($x$) here is the well known error function. 
This probability is of course also that for the $q\bar q'$ to form or
to remain in a bound state. On the other hand, the complementary probability 
of color screening preventing the confinement of a $q\bar q'$ or the melting 
of a hadron would be
\be     P (b > l_D(\t)) =  \int^\infty_{l_D(\t)} \dd b \; F_h(b) 
                        = \text{Erfc} \; (l_D(\t)/a_h \sqrt{\p})
                        = 1 - \text{Erf} \; (l_D(\t)/a_h \sqrt{\p})   \; .
\ee
Note that $\lim_{\,l_D \ra \infty} P(b <l_D) \lra 1$ and 
$\lim_{\,l_D \ra 0} P(b <l_D) \lra 0$.

By taking color screening effects into consideration, the time evolution
equation will have to be modified once again. The color screened parton 
equation now becomes
\be {{\dd f_i^{\text{p}}} \over {\dd \t}} \, \Big |_{p_z \t =\text{constant}} 
    \; = - \; { {f_i^{\text{p}}-f^{\text{p}}_{i\; eq}}
            \over {\q_i^{\text{p}}} }
    \; =   \; \Big (C_i^{\text{p}} - C^{\text{p}}_{i\;q_a\bar q_a'} \Big )
            + C'^{\,\text{p}}_{i\;q_a\bar q_a'}
            + C'^{\,\text{p}}_{i\; g\ra q} + C'^{\,\text{p}}_{i\; p\ra h}
            + C'^{\,\text{p}}_{i\; h\ra p} \; .
\label{eq:cs_p_ev_eq}
\ee
Here $C^{\text{p}}_{i\;q_a\bar q_a'}$ denotes that part of the original
parton collision terms that describes quark $q$ and antiquark $\bar q'$ 
scattering with the pair in a color singlet. Obviously, we have
\be    C^{\text{p}}_{g\;q_a\bar q_a'} = 0 \; .
\ee
The other terms are now weighed by a probability factor and are given by 
\bea   C'^{\,\text{p}}_{i\;q_a\bar q_a'} &=& 
                P(b > l_D) \; C^{\text{p}}_{i\;q_a\bar q_a'} \; ,
\label{eq:coll_noconf}                                            \\
       C'^{\,\text{p}}_{i\; g\ra q}      &=&
                P(b < l_D) \; C^{\text{p}}_{i\; g\ra q}      \; ,
\label{eq:gs_conf}                                                \\
       C'^{\,\text{p}}_{i\; p\ra h}      &=& 
                P(b < l_D) \; C^{\text{p}}_{i\; p\ra h}      \; ,  
\label{eq:coll_conf}                                              
\eea
so that when the screening length becomes very large eventually, all
color singlet $q\bar q'$ pairs would form hadrons via the term represented by
\eref{eq:coll_conf} rather than undergo a scattering through the term
represented by \eref{eq:coll_noconf}. The weight $P(b < l_D)$ 
in \eref{eq:coll_conf} is for the intermediate cluster resonance. 
For the non-perturbative gluon splitting term in \eref{eq:gs_conf}, 
we assumed that it should only be able to function with 
its full strength if the confining forces, which should be the physical origin 
of this term, are unhindered by color screening. So a factor of $P(b < l_D)$ 
is given to this term. Note that the time dependence of $l_D$ has been
suppressed from the above expressions and will not be written out explicitly
below but it should be implicitly understood that $l_D$, $P(b<l_D)$ and
$P(b>l_D)$ are all time dependent quantities naturally. 

The last term in \eref{eq:cs_p_ev_eq} is new and, as far as we are aware,
has not been considered in any parton based models. It is given again by 
a probability weighed term
\be       C'^{\,\text{p}}_{i\; h\ra p} =
                P(b > l_D) \; C^{\text{p}}_{i\; h\ra p}  \; .
\ee
It is there to allow for the possibility that any already formed hadrons 
to have a chance to dissolve back into partons. One can think of this as 
the internal motion of the fermion pair might lead them to wander just a 
bit too far from each other into region where the color force is weakened 
by screening. The dissociation of the hadron would then result. 
The explicit form of $C^{\text{p}}_{i\; h\ra p}$ is
\bea  C^{\text{p}}_{q\; h\ra p} &=& \frac{1}{2 k^0_{h}} \; \frac{S(\a_s)}{2N}
         \;\sum_{q=u,d,s} 
         \int \intkh \intpqb (2\p)^4 \d^{(4)} (k_h -p_q -p_{\bar q}) \nonumber \\
      & & \mbox{\hspace{3.0cm}} \times       
            f_h (1-f_q) (1-f_{\bar q}) \;  |\lan q\bar q|\ch| h\ran |^2  \; ,  \\
      C^{\text{p}}_{g\; h\ra p} &=&  0       \; .
\label{eq:q_diss}
\eea

The collision terms of the color screened time evolution equations for hadrons 
are likewise weighed by similar probabilities. The hadron equations are now 
given by
\be {{\dd f_i^{\text{h}}} \over {\dd \t}} \, \Big |_{p_z \t =\text{constant}} 
    \; = - \; { {f_i^{\text{h}}-f^{\text{h}}_{i\; eq}}
            \over {\q_i^{\text{h}}} }
    \; =   \; C'^{\,\text{h}}_{i\;p \ra h} 
            + C'^{\,\text{h}}_{i\; h\ra p} \; .
\label{eq:cs_h_ev_eq}
\ee
The probability weighed collision terms are then
\bea   C'^{\,\text{h}}_{i\;p \ra h} &=& 
                P(b < l_D) \; C^{\text{h}}_{i\;p \ra h}   \; ,         \\  
       C'^{\,\text{h}}_{i\; h\ra p} &=&
                P(b > l_D) \; C^{\text{h}}_{i\; h\ra p}   \; .
\eea
The last term is similar to \eref{eq:q_diss} above and is given by
\bea  C^{\text{h}}_{h_1\; h\ra p} &=&-\frac{1}{2 k^0_{h_1}} \;S(\a_s) \;\sum_{q=u,d,s} 
         \int \intpq \intpqb (2\p)^4 \d^{(4)} (k_{h_1} -p_q -p_{\bar q}) \nonumber \\
      & & \mbox{\hspace{3.0cm}} \times       
            f_{h_1} (1-f_q) (1-f_{\bar q}) \;  |\lan q\bar q|\ch| h\ran |^2  \; .
\label{eq:h_diss}
\eea
That completes our full set of equations.

\section{Color screening effects on Parton-Hadron Conversion in a Parton Plasma}
\label{sec:results}

In the previous sections, a set of equations were given which describe the
time evolution of a parton plasma ending in a free streaming hadron gas.
Before we could solve for the distributions, we must know the various
transition matrix elements or their squared modulus. The latters should
in general be functions of the kinematic invariants and may also
depend on the coupling. Their nature is clearly non-perturbative. 
To actually know these functions requires a better understanding of
confinement. Since no theory of confinement is available for immediate 
practical use, there is no way of knowing the exact form of these functions. 
Fortunately, our purpose here is to show the effects of the medium
on the parton-hadron transition. This means we can concentrate on the
qualitative differences of the resulting hadron gas coming from the
hadronization of the parton plasma with and without taking the medium into
full account. To do this, we merely have to pick some arbitrary functions that 
give reasonable and sensible results provided that the same functions are
used in both color screened and unscreened hadronization. Different choices
will only affect the rate of the conversion.
The rates themselves are likely to be too hard to obtain experimentally.
For the purpose of our investigation, we have to be satisfied with this approach. 
We thus make the convenient choice of 
\bea   \lan  I |\ch|  I \ran & \sim & \k_I \; \a_s^2                \\
       \lan  q\bar q |\ch|  g \ran & \sim & \k_{g\ra p} \; \a_s     \\
       \lan  q\bar q'|\ch|  h \ran & \sim & \k_{h\ra p} \; ,
\eea
where $\k$'s are constants. The coupling dependence is to give the
rate some strength dependency in order to fulfill the expectation that the 
conversion should be more active in the latter stages. We stress that
this is just a convenient choice and we give no physical justification
for their forms other than requiring the numerics to be not too 
unreasonable. The power of $\a_s$ could for instance be higher or the 
expressions could be polynomials in the coupling in general or there could be 
some other non-analytic expression in $\a_s$. Since there is little knowledge
to indicate why one form is better than an other, we proceed by making
a fairly simple choice. When we compare the screened and unscreened
results, which is our main goal, this choice and the details associated
with it are not so important provided this is fixed for both cases. 
With little information to go on, this is the best we can do for now.

When we investigated the equilibration of the parton plasma, 
the complexity of the collision terms in general meant that only the 
leading order expressions have been used. To be consistent,
the coupling $\a_s$ used in \cite{wong3,wong1,wong2} was also the
one-loop expression even though $\a_s$ was allowed to evolve in time.
Non-perturbative terms are now included in the rate equations 
in the previous sections to make sense of the large coupling region. 
Although we still keep the parton interaction terms when confinement 
is switched on, the emphasis will be on the non-perturbative ones. 
As the equations now bridge two phases, it is perhaps better
to use an $\a_s$ calculated to next-to-leading order or even beyond
that. However in order to be able to compare with the results of earlier
investigations, the leading $\a_s$ expression will be used throughout.
In any case, using the next-to-leading order expression, for example,
will only shift the numerical results up or down but the qualitative
picture will remain intact. 

In the course of the time evolution, the expansion and particle
creation lead to a progressively stronger and stronger interacting parton 
plasma and therefore the use of the perturbative $\a_s$ will eventually
run into the problem associated with the Landau pole. There are various
ways to overcome this and there is not a lack of models of $\a_s$
that are free of this problem. One can see for example any one
of those proposed in \cite{grun,dkt,ss,web2}. The simplest way we can think
of, however, is to freeze the coupling at some point when it becomes 
large. In this way, we could still compare with previous results on the
possible ``back reactions'' of hadronization on the time evolution of the
parton plasma. 

After discussing how we dealt with some of the technical issues, 
we now turn to the results. First we examine the hadron gas converted 
from the parton plasma and then turn to effects on and properties of the
the plasma itself. In \fref{f:h_den}, we plot the variation of the hadron 
number and energy densities with time. The solid curves are the results with
color screening and the dashed ones are without screening. From top to 
bottom in each plot, the curves are for charged pions, neutral pions,
and charged or neutral kaons. As can be seen, without color screening
the growth of the hadron densities is earlier than in the case that the
color screening acts as a barrier to hadronization. The maximum densities
reached are also lower in the color screened hadronization because
the system is under continual expansion. Therefore the hadronic background
to the proposed signatures such as photons, dileptons, strangeness or any
comovers as another possible mechanism for $J/\j$ suppression
\cite{ms} should be less in the real situations when color 
screening is active. It follows that the hadronization time scale would be 
incorrect for any parton based dynamical models that ignore the effects of 
color screening completely. 

\bfi
\centerline{
\epsfig{figure=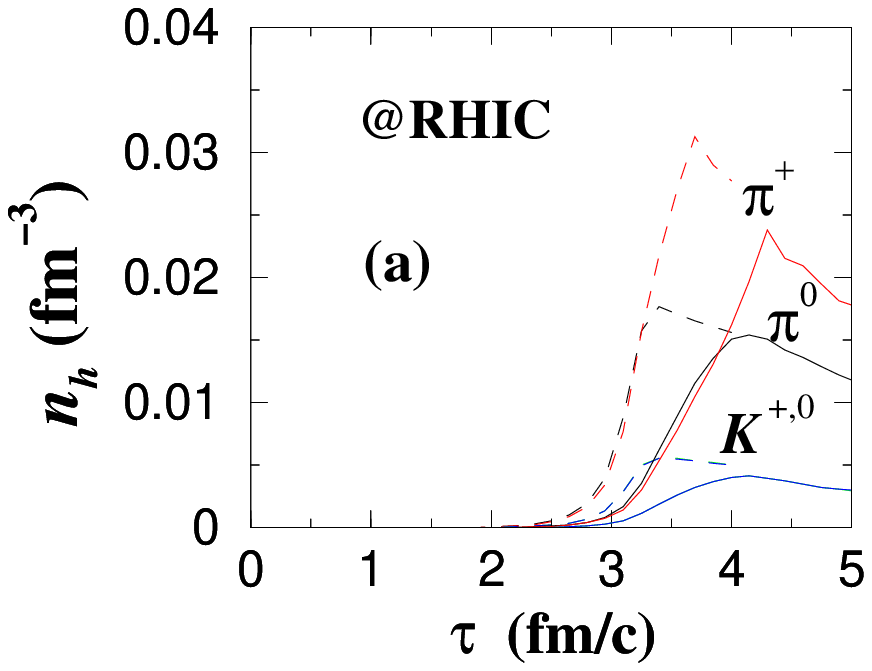,width=2.60in} \hspace{1.50cm}
\epsfig{figure=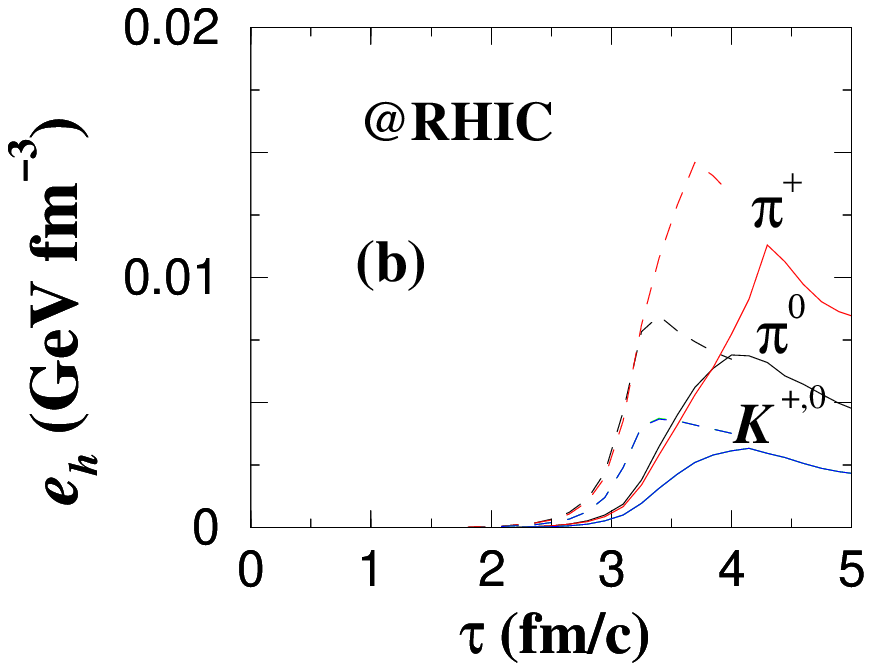,width=2.60in}
}
\caption{The time development of the hadron (a) number and (b) energy densities
at RHIC. The solid (dashed) lines are for hadronization with (without) the barrier
of color screening. The growth of the hadron densities is delayed in the former
case and the maximum densities are lower.}
\label{f:h_den}
\efi

Next we would like to lightly touch upon strangeness which is of great interest
as it is one of the more promising signatures for the quark-gluon plasma.
A great many number of papers have been written on the subject ever since 
it was first proposed in \cite{mr,msm}. One of the more interesting quantities
is the strange to non-strange hadron ratio. It has been argued that the
strangeness saturation parameter $\g_s$ determined from this ratio
gives $\g_s \sim 0.5$ in p-p and $\g_s \sim 0.7$ according to
\cite{soll&etal} in A-A collisions. The latter is still under debate
as refs. \cite{munz&etal1,munz&etal2} found that it was possible to fit AGS
and SPS data with $\g_s =1$, while others such as refs. \cite{soll&etal,jj} 
disagreed. A marked difference in $\g_s$ potentially suggests a quite 
different scenario in the collision core, one that may support a
much more efficient production of strangeness. With our qualitative model, 
we do not attempt to make any statement about the definite value of 
$\g_s$ in A-A collisions. For that, one needs a good quantitative model. 
Here we merely make the observation that in our model, in which we do not 
include hadronic re-scattering, the strange to non-strange ratio progressively 
lowers down to the expected level from a rather high value to somewhere 
above 0.1 in \fref{f:sns}. We attribute this to the progressive reduction of 
phase space for kaons since the probability of the formation of the more 
massive color singlet clusters with invariant mass $M > m_\p+m_K$ is less as
time progresses and the hadronization part of our model does not bias against 
the strange flavor. With and without color screening, there are some
differences in this ratio which may not be so easy to see on the plot. 
The general behavior is the same in both cases, except in the case without 
color screening the curve ends up somewhat higher than the color screened 
one. The eventual strange to non-strange ratio therefore tends to be slightly 
lower when color screening is included simply because of phase space. 
\bfi
\centerline{
\epsfig{figure=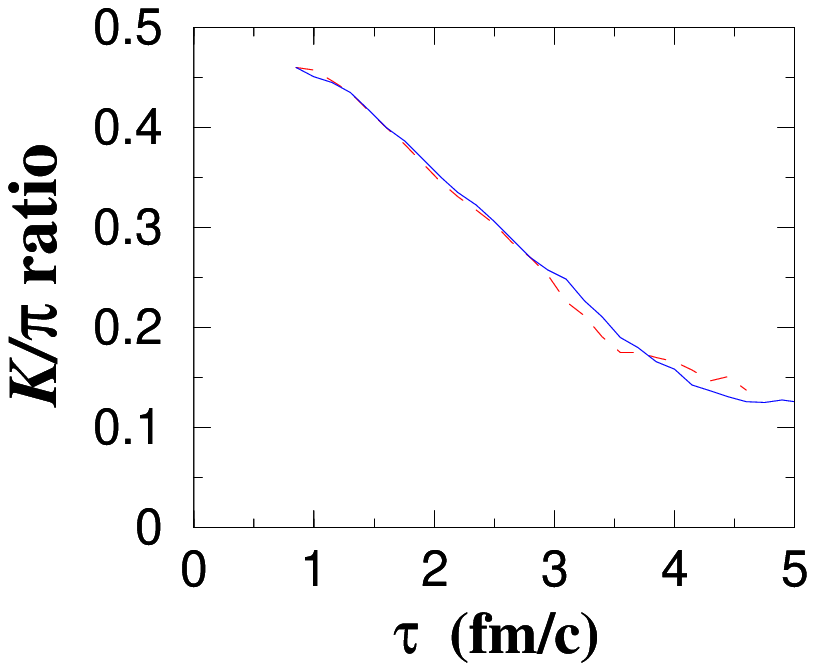,width=2.40in} 
}
\caption{The strange to non-strange hadron ratio as a function of time
with color screening (solid line) and without color screening (dashed line).}
\label{f:sns}
\efi

\bfi
\centerline{
\epsfig{figure=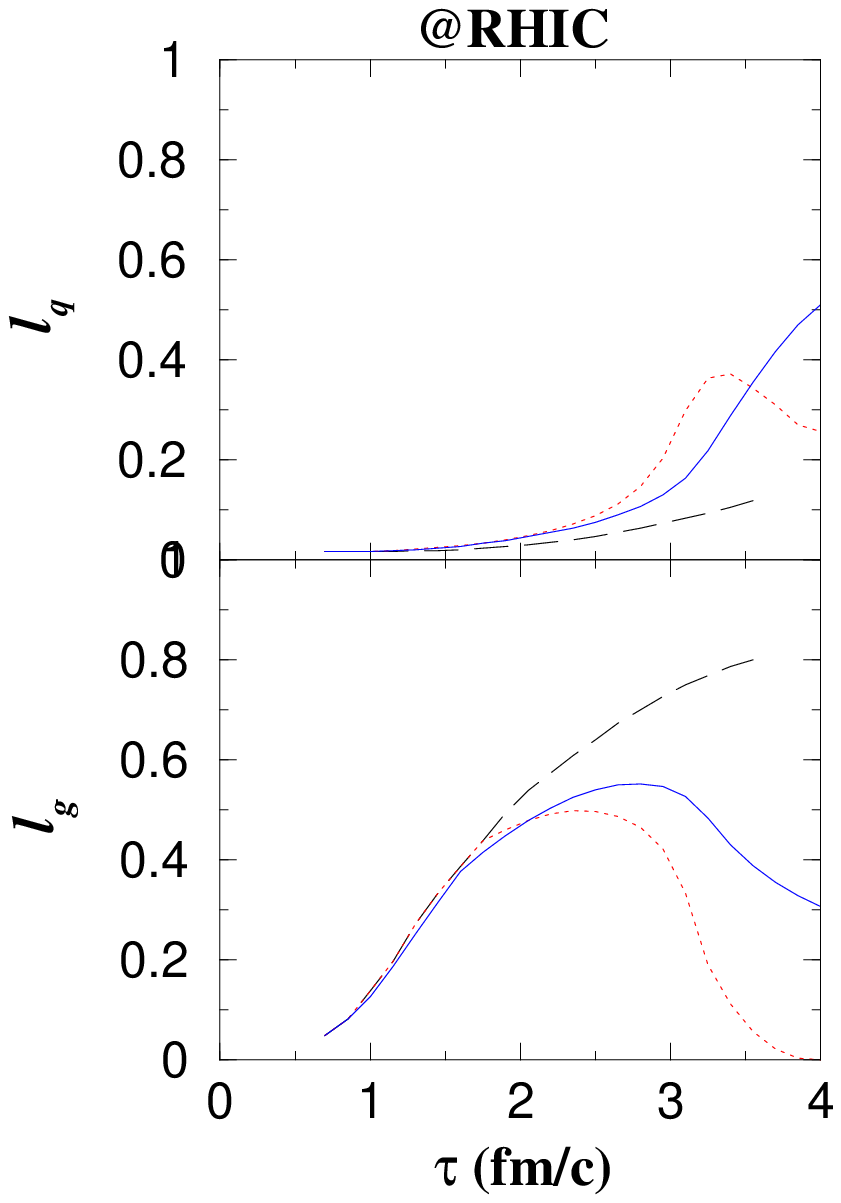,width=2.50in}}
\caption{Chemical or parton equilibration here expressed in terms of the 
gluon and quark fugacities is seen to be disrupted by the hadronization
independent of whether there is color screening (solid line) or there 
is no color screening (dotted line). The original case (long dashed line),
where hadronization was not considered, is also shown for comparison.} 
\label{f:p_fug}
\efi

We now turn to the effect of hadronization on the parton plasma.
In \fref{f:p_fug}, we plot the quark $l_q$ and gluon $l_g$ fugacities. 
Three cases are presented in the plot. The original time evolution
of the parton plasma without hadronization are shown in long
dashed lines. Then of the two cases with hadronization, dotted lines
are for the case with no color screening and the solid lines are
with color screening. The gluon fugacity $l_g$ which rose to above 
0.8 in the original case is now affected quite significantly by
the hadronization process. Instead of rising continuously, 
the two $l_g$ curves with hadronization now rise first,
then stop ascending altogether before reversing to a downward 
descent. Remembering that we have included a term for non-perturbative
gluon splitting which is prerequisite for hadronization into hadrons
consisting only of quark and antiquark, this term is responsible
for preventing $l_g$ from retracing the old ascending path. On the
other hand, this same term is now enhancing the rise of $l_q$. 
We see that they are given a boost in the equilibration, rising at
maximum to three or four times above the curve without hadronization. 
Confinement therefore interrupts the otherwise smooth time evolution
of the partons. Comparing the two cases with hadronization, the one
with color screening is again delayed as manifested already in the
hadron density plots. At RHIC, it is highly unlikely that parton
equilibration can be completed before the phase transition.
By including hadronization, we show here that it is even more 
unlikely.

\bfi
\centerline{
\epsfig{figure=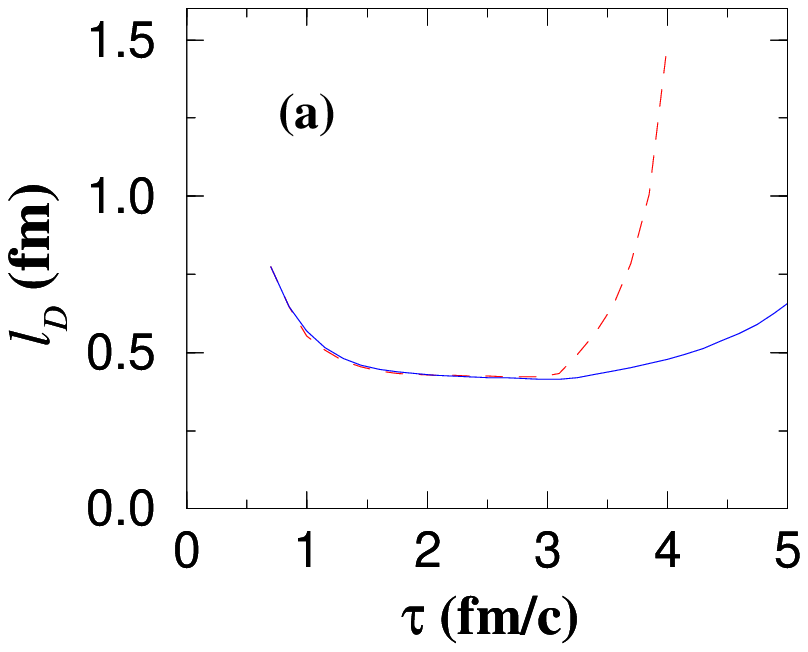,width=2.50in} \hspace{1.50cm}
\epsfig{figure=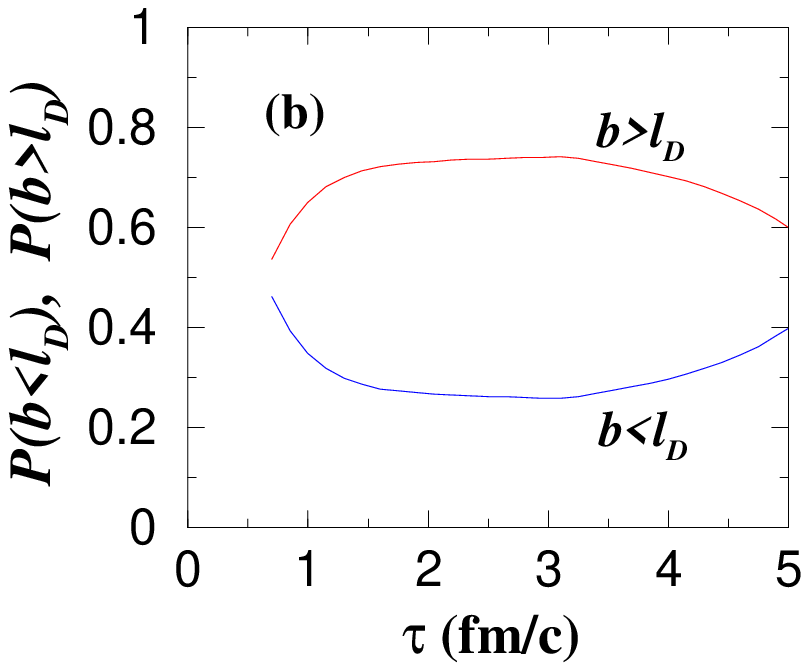,width=2.50in} }
\caption{(a) The variation of the screening length $l_D$ during the time 
evolution ending in hadronization with (without) color screening is
shown in solid (dashed) line. The two curves deviate from each other
significantly at later times when hadronization is well underway.
(b) The corresponding probabilities $P(b>l_D(\t))$ and $P(b<l_D(\t))$ as a 
function of time for the case with screening.} 
\label{f:ld}
\efi

Lastly to argue for the importance of including both confinement 
and color screening, and to gain some insight into the internal 
struggle between the two, we show the variation of the screening 
length $l_D$ and the associated probabilities $P(b>l_D)$, $P(b<l_D)$ 
with $\t$ in \fref{f:ld}. At the time around $\t=3$ fm, confinement has 
essentially been turned on. One can see that there is a huge difference
in \fref{f:ld} (a) between the two cases of screening (solid line) and no
screening (dashed line). In the latter, hadronization proceeded in an
unhindered fashion therefore the parton medium essential for
weakening the confining forces was quickly depleted. As a result the
screening distance grew quickly. When the color screening barrier was 
erected against hadronization, the parton medium was able to hold itself
together and maintained its screening power for much longer. In this 
latter case, the associated probabilities $P(b>l_D)$ and $P(b<l_D)$
are plotted in \fref{f:ld} (b). The much slower rise of $l_D$ meant 
that the probabilities also varied slowly although they had the 
expected behavior at larger times. We see trace of the one-loop behavior
of the screening mass manifesting here too. Since it is rather
time consuming to evolve the system on a computer so we stopped the 
time evolution once the qualitative features could be extracted. 
It is expected that eventually $P(b>l_D)$ and $P(b<l_D)$ will reach
their correct limits at very large times. Relating \fref{f:h_den}
and \fref{f:ld}, one might perhaps expect the value of the probability
$P(b<l_D)$ should be larger or equivalently $P(b>l_D)$ smaller at the 
moment when the maximum densities in \fref{f:h_den} occurred. This 
seemingly straight forward expectation is, however, not entirely
justified because these maxima are results of a combination of
factors: the actual sizes of the transition matrix elements, 
the switching factor $S(\a_s)$ discussed in Sec. \ref{sec:part-had} and 
the available parton densities. Confinement might be more favorable at 
later times but the number of available partons or the parton densities 
would be lower. 

Finally, we would like to briefly discuss the problem with the 
transition matrix elements. We have already mentioned that there 
is no easy way to get readily usable expressions for these elements. 
Our chosen form is for convenience and it fulfills at least some 
aspects of the physical expectations. As to the constants, to obtain 
reasonable sensible results, there is a range of possible values. 
For the actual confinement matrix elements, we have actually chosen 
$\k_0 = \k_1 = \k_{1/2}$ for simplicity, since our main concern was 
the color screening effects and to point out that this aspect had 
been missing in the existing models. Using different values of these 
constants will change the actual numbers but the shape of the plots, 
the time sequence and therefore the qualitative picture that we tried 
to extract will remain intact.

\section*{Acknowledgments}

The author would like to thank Joe Kapusta for useful discussion,
and Paul Ellis for critical reading of the manuscript. 
This work was supported by the U.S. Department of Energy under 
grant DE-FG02-87ER40328.

\end{document}